\documentclass[aps,pra,twocolumn,10pt,superscriptaddress,showpacs,nofootinbib,longbibliography
]{revtex4-1}
\usepackage{hyperref,xcolor}

\newcommand{\QUOTE}[2]{\begin{itemize}\item[]\emph{``#1"}  \\ \rule{0pt}{0pt} \hfill (#2) \end{itemize}}

\newcommand{\tabprotocol}{
\begin{table*}
\caption{\label{tab:protocol}Interview questions, presented in the order the were asked during the interview.}
\begin{ruledtabular}
\begin{tabular}{p{0.005\textwidth}p{0.95\textwidth}}
\multicolumn{2}{l}{Departmental and course context} \\
1. & How many majors graduate from your department each year?\\
2. & Including electronics courses, how many total lab courses beyond the first year are offered in your department?\\
3. & How many are focused on electronics?\\
4. & How many students typically enroll in the electronics course each term?\\
5. & How many sections are offered?\\
6. & How many instructors, teaching assistances, and/or learning assistants per section?\\
7. & Do students typically work alone, in pairs, or in groups?\\
8. & What topics do you typically cover?\\
9. & Is there a final project associated with the course?\\
10. & Is there anything special about this course that you want to tell me about?\\ \\
\multicolumn{2}{l}{Teaching history} \\
11. & Have you previously taught the electronics course? If so, how many times?\\
12. & Are you currently teaching the electronics course?\\ \\
\multicolumn{2}{l}{Relevance of troubleshooting} \\
13. & What do you think is the purpose of electronics courses?\\
14. & I'm interested in the role of troubleshooting in electronics courses. Before we dive into this, can you tell me what you think ÒtroubleshootingÓ means in the context of electronics?\\
15. & To what extent is the ability to troubleshoot related to the educational purpose of electronics courses, if at all?\\
16. & What about more generally? To what extent is the ability to troubleshoot important beyond the context of electronics courses? \\ \\
\multicolumn{2}{l}{Teaching and assessment practices} \\
17. & When students encounter problems in electronics labs, how do they typically respond? For example, do they try to figure it out on their own, do they immediately ask for help from classmates or the instructor, or do they do something else?\\
18. & What strategies do students typically try before asking for help, if any?\\
19. & What strategies do you want them to have tried before asking for help, if any?\\
20. & When you help students troubleshoot problems, what do you usually say or do?\\
21. & How do you know if students are good at troubleshooting?\\
22. & How do you know if students are bad at troubleshooting?\\
23. & Have you ever designed an activity to test students' troubleshooting skills? If so, what did you do and how did it go?\\
24. & In what ways do you teach about troubleshooting, if at all?\\
25. & Have you ever implemented activities that were specifically designed to improve students' ability to troubleshoot? If so, what did you do and how did it go?\\
26. & Have you ever talked about troubleshooting in your lectures? If so, what did you say and how did it go?\\
27. & Have you ever addressed troubleshooting in any curricular materials such as the syllabus, lab guides, or handouts? If so, what did you do and how did it go?\\
28. & Do you think people can learn to troubleshoot or is it an innate skill?\\ \\
\multicolumn{2}{l}{Wrap-up questions} \\
29. & When it comes to teaching or evaluating troubleshooting in electronics courses, what kind of materials, activities, or other resources would you be interested in, if any?\\
30. & Is there anything else you'd like to tell me?\\
31. & I'm trying to reach out to a broad range of instructors from different institution types to be sure I collect diverse perspectives about electronics and troubleshooting. I already have institutional information from the Carnegie classification system. For the transcript record, is it okay if I ask about your gender and race?
\end{tabular}
\end{ruledtabular}
\end{table*}
}

\begin{document}


\title{{Electronics lab instructors' approaches to troubleshooting instruction}}

\author{Dimitri R. Dounas-Frazer}
\email{dimitri.dounasfrazer@colorado.edu}
\affiliation{Department of Physics, University of Colorado Boulder, Boulder, CO 80309, USA}

\author{H. J. Lewandowski}
\affiliation{Department of Physics, University of Colorado Boulder, Boulder, CO 80309, USA}
\affiliation{JILA, National Institute of Standards and Technology and University of Colorado Boulder, Boulder, CO 80309, USA}

\date{\today}

\begin{abstract}
In this exploratory qualitative study, we describe instructors' self-reported practices for teaching and assessing students' ability to troubleshoot in electronics lab courses. We collected audio data from interviews with 20 electronics instructors from 18 institutions that varied by size, selectivity, and other factors. In addition to describing participants' instructional practices, we characterize their perceptions about the role of troubleshooting in electronics, the importance of the ability to troubleshoot more generally, and what it means for students to be competent troubleshooters. One major finding of this work is that, while almost all instructors in our study said that troubleshooting is an important learning outcome for students in electronics lab courses, only half of instructors said they directly assessed students' ability to troubleshoot. Based on our findings, we argue that there is a need for research-based instructional materials that attend to both cognitive and non-cognitive aspects of troubleshooting proficiency. We also identify several areas for future investigation related to troubleshooting instruction in electronics lab courses.
\end{abstract}

\maketitle


\section{Introduction}\label{sec:intro}

Troubleshooting is a critical skill in science, technology, engineering, and mathematics disciplines~\cite{Perez1991,Johnson1995,Jonassen2006}, including experimental physics~\cite{Pollard2014,Holmes2016}. Accordingly, the ability to troubleshoot is an important design-related learning outcome for undergraduate physics laboratory courses~\cite{AAPT2015,Zwickl2012,Zwickl2013}. {In particular}, electronics courses are ideal environments for physics students to practice and hone their troubleshooting skills because students naturally engage in troubleshooting during most circuit-building lab activities. However, troubleshooting is only one of many potential foci for electronics courses~\cite{AAPT2015}, and it is unclear to what extent electronics instructors emphasize this skill in their course goals or their teaching and assessment practices. {Understanding instructors' perceptions of troubleshooting and its connection to electronics courses is necessary for the development of research-based assessments and activities that are relevant to instructors. In addition, understanding instructors' teaching and assessment practices could infuse future transformation efforts with creative ideas already being implemented by seasoned practitioners. To these ends, we report on instructors' perceptions about, and experiences with, troubleshooting instruction in electronics courses.}

Prior work on troubleshooting in electronics courses has focused on students' actions, interactions, and learning. At the undergraduate level, we have previously studied the role of model-based reasoning~\cite{Dounas-Frazer2016a,Dounas-Frazer2015} and socially-mediated metacognition~\cite{VanDeBogart2015} in physics students' approaches to troubleshooting electric circuits. Other work has focused on developing training programs for students in engineering~\cite{Schaafstal2000,deCroock1998} and technical~\cite{Johnson1993,Johnson1999,MacPherson1998,Ross2007,Pate2011} fields. At the high school level, past studies involved identifying expertise-related differences among students troubleshooting simulated circuits~\cite{vanGog2005a, vanGog2005b} and evaluating the effectiveness of various instructional strategies~\cite{vanGog2008, vanGog2006, Kester2006, Kester2004} on students' troubleshooting performance. Other work in electronics courses has focused on the design~\cite{Halstead2016,Lewandowski2015,Shaffer1992} and evaluation~\cite{Coppens2016b,Mazzolini2011,Getty2009} of courses for physics and engineering students, as well as on student understanding of electric circuits~\cite{Engelhardt2004,Papanikolaou2015,Coppens2012} and electronics concepts~\cite{Stetzer2013,McDermott1992,*McDermott1993}.

Studies on instructor beliefs and practices can complement student-focused research to produce a more complete understanding of a particular learning environment. For example, understanding  instructors' perspectives can clarify the need for, and objectives of, research-based assessments, {two important aspects of assessment development}~\cite{Engelhardt2009,Wilcox2015}. While there are many studies on the views and practices of physics educators~\cite{Dancy2007,Henderson2007,Yerushalmi2007,Lasry2014,Turpen2016,Goertzen2010,Maries2016,Spike2016}, these studies tend to focus on instructors~\cite{Dancy2007,Henderson2007,Yerushalmi2007,Lasry2014,Turpen2016} or teaching assistants~\cite{Goertzen2010,Maries2016,Spike2016} in introductory lecture, studio, or tutorial undergraduate environments---not labs. In the context of engineering education, some work has focused on thermodynamics, electronics, and statics instructors' pedagogical beliefs in upper-division theory-based contexts~\cite{Borrego2013} and electronics instructors' learning goals for labs~\cite{Coppens2016a}. We are unaware of prior research that focuses on electronics lab instructors' teaching approaches.

We explore electronics lab instructors' goals and practices related to teaching and assessing troubleshooting.  {Along these lines, we identified six research questions:}
\begin{enumerate}
\item[RQ1.] {According to instructors, what is the purpose of teaching electronics lab courses?}
\item[RQ2.] Do electronics instructors think it is important for students to be able to troubleshoot and, if so, why?
\item[RQ3.] How do electronics instructors define troubleshooting in the context of the electronics lab?
\item[RQ4.] How do electronics instructors characterize troubleshooting proficiency?
\item[RQ5.] How do electronics instructors describe their approaches to teaching students how to troubleshoot?
\item[RQ6.] How do electronics instructors describe their approaches to assessing students' ability to troubleshoot?
\end{enumerate}
{This work not only helps clarify whether and how physics education researchers might move forward in the realm of troubleshooting instruction in electronics courses, it is a first step towards understanding physics lab instructors' {instructional beliefs and practices} more generally.} We report on interviews with 20 electronics instructors. Preliminary results from this study have been reported elsewhere~\cite{Dounas-Frazer2016b}. Here we provide a more comprehensive analysis, organized as follows.  In Sec.~\ref{sec:background}, we {define troubleshooting as a cognitive task and provide a synopsis of work on teaching strategies relevant to troubleshooting in electronics}. In Sec.~\ref{sec:methods}, we describe our study methods and participant population. We present the results of our analyses in Sec.~\ref{sec:results}. Finally, in Sec.~\ref{sec:discussion}, we summarize our work, discuss implications of our findings, and highlight potential avenues for future study.


\section{Background}\label{sec:background}

In this section, {we familiarize the reader with the language and ideas that are common in discussions of troubleshooting competence and instruction. To that end, we describe troubleshooting as a cognitive task and we provide an overview of cognitive apprenticeship as it relates to troubleshooting instruction}. In doing so, we define several terms related to the doing and teaching of troubleshooting {that we use when presenting the results of our study}.


\subsection{Troubleshooting as a cognitive task}\label{sec:troubleshooting}

Troubleshooting is the process of diagnosing and repairing a malfunctioning system. Cognitive task analyses of troubleshooting typically describe the types of knowledge, cognitive subtasks, and strategies required for competent troubleshooting~\cite{Jonassen2006,Schaafstal2000,Johnson1988}. We have summarized the cognitive aspects of troubleshooting electric circuits comprehensively and in detail elsewhere~\cite{Dounas-Frazer2016a}. Here, we focus on the knowledge, subtasks, and strategies that are most relevant to the present work.

Competent troubleshooting is facilitated by multiple types of knowledge, including \emph{domain}, \emph{system}, \emph{procedural}, and \emph{strategic} knowledge. Domain knowledge consists of the theories and concepts underlying the system being troubleshot, including models like Ohm's law and concepts like conservation of charge. System knowledge involves understanding the structure and function of a system, including recognizing that a complex circuit is made up of multiple interacting subsystems. Procedural knowledge refers to the appropriate use of test and measurement equipment, such as function generators and oscilloscopes. Lastly, strategic knowledge refers to the heuristic techniques and methodical approaches to troubleshooting the system.

In addition to requiring mastery of multiple types of knowledge, the process of troubleshooting can be subdivided into multiple subtasks, including \emph{generating causes} and \emph{performing tests}. {Generating causes} involves formulating causal hypotheses about {potential} sources of malfunction, and {performing tests} involves performing diagnostic measurements to determine whether or not a proposed fault is an actual fault. During the process of troubleshooting, these tasks are carried out in iterative and nonlinear ways. For example, if testing reveals that none of the proposed faults is an actual fault, then the troubleshooter must generate additional causes.

{To navigate the recursive troubleshooting process, troubleshooters rely on various strategies. While myriad troubleshooting strategies exist~\cite{Jonassen2006}, two strategies for troubleshooting electric circuits were commonly discussed by participants in our study: the \emph{forward topographic strategy} and the \emph{split-half strategy}.} The {forward topographic strategy} refers to the process of making measurements starting at the input and working towards the output along a ``pathway" in the circuit (e.g., the flow of electrons or a sequence of voltage drops). In this manner, the troubleshooter engages in a quasi-linear search for faults. The {split-half strategy}, on the other hand, is a way to iteratively reduce the problem space through a binary search: the troubleshooter divides the circuit into two subsystems, performs a diagnostic measurement at the midpoint, isolates one of the two subsystems as a source of fault, and repeats the process in that subsystem until the faulty component or connection has been identified.

Cognitive task analyses do not fully capture the process of troubleshooting; non-cognitive aspects are also important. For example, Estrada and Atwood~\cite{Estrada2012} identified troubleshooting as the most common source of frustration for students in introductory physics labs, and MacPherson~\cite{MacPherson1998} noted that ``personality characteristics such as perseverance, ingenuity, self confidence, and patience are common to both technological troubleshooting and general problem-solving." In the present work, we will show that the instructors in our study described both cognitive and non-cognitive aspects of troubleshooting proficiency.


\subsection{Troubleshooting instruction and apprenticeship}

Farnham-Diggory~\cite{Farnham-Diggory1994} identified three paradigms of instruction: the \emph{behavior}, \emph{development}, and \emph{apprenticeship} models. Each model is characterized by two factors: the distinction between novices and experts and the mechanism by which novices become experts. In the behavior model, novices are distinguished from experts by their relatively low performance on quantitative measures of proficiency, and novices become experts through accumulation of skills and knowledge that result in incrementally higher (\emph{i.e.}, more ``expert-like") scores on these measures. In the development model, novices and experts have different qualitative models of phenomena; novices' models are questioned, challenged, and otherwise perturbed, ultimately pushing novices to revise their ways of thinking in qualitative ways. Finally, in the apprenticeship model, novices and experts differ culturally and experts transmit tacit, context-dependent knowledge to novices through acculturation.

Similar to literature in other domains~\cite{Farnham-Diggory1994}, articles about troubleshooting instruction~\cite{Johnson1995,Schaafstal2000,deCroock1998,Johnson1993,MacPherson1998,Ross2007,Pate2011, vanGog2008, vanGog2006, Kester2006, Kester2004} seldom specify the paradigm of instruction underlying their curricular designs. Nevertheless, some previous work on troubleshooting instruction~\cite{Jonassen2006,Johnson1999} explicitly employs a form of apprenticeship, namely, \emph{cognitive apprenticeship}. Cognitive apprenticeship is a type of apprenticeship where the focus of the learning experience is on cognitive and metacognitive skills and processes~\cite{Collins1991,Hennessy1993}. Whereas observation of novices by experts (and vice versa) as they work alongside one another plays a crucial role in traditional apprenticeship models, cognitive apprenticeship requires novices and experts alike to make their thinking visible, often by verbalizing their thought processes out loud. Because troubleshooting is, in part, a cognitive task, cognitive apprenticeship is salient to troubleshooting instruction.

Cognitive apprenticeship involves multiple teaching methods, including \emph{modeling}, \emph{coaching}, \emph{scaffolding}, \emph{fading}, and \emph{articulation}~\cite{Collins1991}. Modeling involves an expert troubleshooting circuits while verbalizing their mental processes so that students can observe and listen. Coaching refers to the expert observing students troubleshoot and offering hints, tips, and reminders. Scaffolding and fading involve experts providing suggestions and help early on, but gradually removing supports as time passes. Articulation is the process of getting students to verbalize their understanding of the problem, their hypotheses, or their troubleshooting strategies.

Cognitive apprenticeship also involves cooperative problem solving and a gradual increase in the complexity of tasks that students are asked to complete~\cite{Collins1991}. These features are consistent with the typical electronics lab course: instructors provide individualized support to students, students work in pairs, and circuit-building activities become increasingly complicated as the course progresses. Indeed, as we will show, instructors in our study described many teaching strategies that align with the cognitive apprenticeship model.

One major goal of this paper is to describe electronics instructors' perspectives on, and experiences with, teaching students how to troubleshoot circuits. We focus not only on troubleshooting competence and instruction, but also other facets of instruction (\emph{e.g.}, assessment) that, to our knowledge, have not previously been discussed in the troubleshooting literature. {Therefore, while the results presented in this paper overlap with frameworks for troubleshooting as a cognitive task and troubleshooting instruction as a form of apprenticeship, they are not fully characterized by these frameworks.} In the following section, we describe our study in more detail.


\section{Methods}\label{sec:methods}

Because there have been few investigations of electronics lab instruction, this study is qualitative and exploratory. {We conducted in-depth interviews with electronics lab instructors, focusing on their ideas about, and approaches to, the teaching and assessing of troubleshooting in electronics courses. Our study enables us to describe learning goals and instructional strategies in the words of our interviewees.} These {descriptions} are likely biased in favor of electronics instructors who enjoy engaging in reflective conversations about their teaching. Additionally, we {solicited} self-reported information about teaching practices; we did not perform any observations of electronics lab instruction. As is generally true of exploratory qualitative investigations, our results are not {necessarily} generalizable to the broader population of electronics lab instructors. Nevertheless, despite these limitations, we believe that this study not only provides valuable insight into the role of troubleshooting in electronics lab courses, it also identifies several open questions that can be studied in future work.

{In this section, we describe our participant recruitment efforts, the course context for the instructors in our study, and our methods of data collection and analysis.}


\subsection{Participant recruitment}

One methodological goal of this study was to interview instructors teaching electronics courses at a variety of institution types. To this end, we generated a database of electronics instructors and used the Carnegie classification system~\cite{Carnegie2015} to characterize the institutions with which they were affiliated. Our database was initially populated with instructors from our own professional networks as well as members of the Advanced Laboratory Physics Association~\cite{ALPhA} who participated in conference sessions and workshops related to electronics at the 2015 Conference on Laboratory Instruction Beyond the First Year. In addition, we perused websites for physics departments at Minority-Serving Institutions and Women's Colleges in order to identify whether those departments offered an electronics course. If so, we added the corresponding instructor to our database.

During Fall 2015, we solicited participation from 47 instructors {in our database} via email. {In total, 20 instructors from 18 distinct institutions agreed to participate. A variety of institutions were represented in our study:} 12 public and 6 private not-for-profit institutions; 9 Predominantly White Institutions, 6 Hispanic-Serving Institutions, 2 Women's Colleges, 1 Historically Black College or University, and 1 Tribal College or University. One institution was classified as both Predominantly White and a Women's College. In terms of size and selectivity, instructors from small, medium, and large institutions as well as from inclusive, selective, and more selective institutions were about equally represented in our data set. Three institutions were two-year colleges, 5 were four-year institutions, 8 were Master's-granting institutions, and 3 were universities with doctoral physics programs. 

{Of the 20 participants in our study,} 15 identified as white or Caucasian alone, 2 identified as mixed race (1 white and Black, 1 Caucasian with Cherokee and African background), and 1 each identified as Asian Indian, Mexican American, and Persian. In addition, 14 identified as male and 6 as female. In order to maintain anonymity of research participants, we do not report intersections of race or ethnicity and gender. {These instructors had varying levels of experience teaching electronics:} 7 had taught the course 2--5 times, 5 had taught it 6--10 times, 6 had taught it 11--20 times, and 2 had taught the course more than 30 times. In addition, 10 instructors were actively teaching electronics during {Fall 2015}, when the interviews were being conducted. {Of the {10} instructors who were not actively teaching electronics at the time of the interview, 6 were planning to teach the course the following semester.} In the next subsection, we describe the types of courses taught by the participants in our study.


\subsection{Departmental and course context}

{We did not conduct a comparative study of instructor practices in different types of departments or courses. Nevertheless, we describe the types of departments and courses represented in this study in order to provide the interested reader with more detailed information about the context of our study.} Information about departmental and course context was self-reported by instructors at the start of the interview, as discussed in the following subsection.

For 5 of the 18 institutions in our study, the electronics course discussed during the interview was part of a doctoral physics program, an undergraduate engineering program, or a pre-engineering program at a two-year college. For the other 13 institutions involved in the study, the electronics course discussed during the interview was a required part of the undergraduate physics curriculum.

Of the 13 institutions that included electronics as part of their undergraduate programs, 7 had small physics departments (3 to 10 physics bachelor degrees awarded per year); 4 had medium departments (11 to 20 degrees per year), and 2 had departments that awarded over 30 degrees per year. The size of the electronics course was about equally split across physics departments: there were roughly equal numbers of small (3 to 10 students), medium (11 to 20 students) and large (21 to 30 students) class sizes in our data set. One institution had an enrollment of over 30 students in its electronics course for physics majors. In 10 of the physics departments, there was only 1 undergraduate lab course dedicated to electronics; the other 3 departments offered 2 or 3 such courses. When interviewing instructors from departments with multiple electronics courses, the interview focused on the course most recently taught by the interviewee. 

At 12 of the 18 institutions, each lab section of the electronics course was taught by a single instructor. At the other 6 institutions, the instructor had the support of a teaching or learning assistant. At 13 institutions, students worked in pairs during lab. Students worked individually at 4 institutions and in triplets at 1 institution.  Across all 18 institutions, students typically built both analog and digital circuits. Common circuit types included filters, amplifiers, oscillators, analog-to-digital converters, counters, and flip-flops. Common circuit components included resistors, capacitors, inductors, transistors, diodes, operational amplifiers, and logic gates. In some courses, students also worked with microcontrollers. Students typically used power supplies, oscilloscopes, and multimeters to perform measurements. To help students model and analyze circuits, most instructors  taught Ohm's Law, Kirchhoff's Laws, and node and mesh analysis. Some instructors also covered Laplace transforms, Fourier transforms, and/or basic solid state concepts such as P-N junctions or electron transport. At most institutions, the electronics lab culminated in a final project.

The variety of educational levels and contexts represented in our study is due, in part, to our attempts to solicit input from instructors teaching at a variety of institution types. In this exploratory study, we aim to provide an in-depth picture of the varied practices and beliefs of the electronics instructors who participated in our study.


\subsection{Data collection and analysis}

\tabprotocol

We conducted 20 semi-structured interviews using a protocol that was informed by our research questions. We collaboratively designed the interview protocol so that the questions would provide a comprehensive picture of a particular interviewee's approach to troubleshooting instruction. Our protocol consisted of 31 questions: 10 focused on departmental and course context, 2 on participants' teaching history, 4 on the value of troubleshooting and its connection to the purpose of electronics, 12 on teaching and assessment practices related to troubleshooting, and 3 ``wrap-up" questions including one final question about participants' race, ethnicity, and gender. The interview questions are provided in Table~\ref{tab:protocol}. Most deviations from the protocol were instances where the interviewer asked a participant to clarify or elaborate on an idea.

Interviews were conducted via videoconference or in-person. Audio data were recorded for each interview. Each interview lasted about 35--55 minutes, for a cumulative total of about 15 hours of audio data. One of us (D.R.D.F.) conducted and transcribed all interviews. The transcripts were the data that we analyzed.

{While we have previously used a cognitive task analysis of troubleshooting (Sec.~\ref{sec:troubleshooting}) as an \emph{a priori} coding scheme in a different study~\cite{Dounas-Frazer2016a}, we did not use that or other existing frameworks in a similar capacity in the present work. In this exploratory study, our goal was not to map a particular framework onto instructors' self-reported conceptions or practices. Rather, our goal was to portray the breadth of instructors' perspectives about, and experiences with, troubleshooting instruction. Hence, the frameworks for troubleshooting practice and instruction presented in Sec.~\ref{sec:background} simply provide a vocabulary with which to discuss results, and our analysis is centered around our research questions RQ1--RQ6.}

Six themes, informed by our research questions, served as an an \emph{a priori} coding scheme: (i) the purpose of electronics courses, (ii) the value of troubleshooting, (iii) the definition of troubleshooting, (iv) characteristics of proficient troubleshooting, (v) methods of teaching troubleshooting, and (vi) methods of assessing troubleshooting. For each theme, one of us (D.R.D.F.) read through each transcript and identified related ideas that emerged across interviewees. These ideas were discussed by both authors and collaboratively grouped into subthemes. Each transcript was read in its entirety a total of six times and the authors reached consensus on all subthemes. In the following section, we report the results of this analysis.


\section{Results and interpretation}\label{sec:results}

We describe 20 electronics lab instructors' perspectives on troubleshooting instruction. Our goal is to richly depict the experiences and perspectives of the instructors in our study. Since we are not performing a comparative analysis, we do not distinguish between the varied contexts in which the instructors teach.

We selected excerpts from the transcripts that exemplify the subthemes in our coded data; we do not present extreme cases. We believe it is useful to provide some indication of whether certain ideas or practices were described by many instructors or just a few. Accordingly, we occasionally use the following qualifiers when presenting results: \emph{almost all} (17--19 instructors), \emph{many} (12--16), \emph{about half} (9--11), \emph{some} (4--8),  and \emph{few} (2--3). If an idea was expressed by all 20 participants or by only 1 participant, we say so explicitly. When providing excerpts, we indicate the speaker's pseudonym in parentheses. After presenting excerpts, we often restate them in our own words in order to clarify our interpretation. 

We organize our results such that each of the following subsections corresponds to one of the six research questions presented in Sec.~\ref{sec:intro}.


\subsection{Purpose of electronics}

When asked about the purpose of electronics courses, many participants said that electronics is a gateway to advanced lab coursework, graduate-level research, and/or careers that involve research and development. One participant who taught at a two-year college said that the purpose of the course was to prepare his students to succeed in advanced lab coursework at the university to which most of his physics students transfer. Some instructors said that the course makes their students ``useful in the lab," referring to instructional and/or research lab contexts.

In terms of learning goals, almost all of the instructors in our study said that developing students' ability to troubleshoot is an important learning goal of the electronics course:
\QUOTE{[Troubleshooting is] central. It absolutely is central. In fact, I think it's one of the most important things they learn in electronics.}{Walnut}
\QUOTE{Electronics is a lot about troubleshooting. \ldots\ One of my biggest goals is troubleshooting skills.}{Larch}
\QUOTE{The most important thing for me in the class with the students is that they learn troubleshooting.}{Filbert}
In addition to troubleshooting, participants articulated a variety of other learning goals for the electronics course: developing basic circuit knowledge; learning how to use common test and measurement equipment, like function generators and oscilloscopes; and developing the ability to design, construct, and model circuits. Some instructors said that the course was an opportunity for students to connect theoretical knowledge to real-world systems, while others said that developing students' theoretical knowledge was not a major goal of the course.

Other, less commonly articulated course goals included improving students' collaboration, communication, and time management skills, as well as their confidence, creativity, and independence. Two instructors emphasized the importance of teaching lab safety practices since some students may go on to work in environments with high currents. Lastly, one instructor who taught at a Women's College said that one of her goals was to ``turn out more girl geeks." This goal was facilitated by the single-gender environment in which she taught: ``it's never the case that the men are doing everything and the women aren't."


\subsection{Value of troubleshooting}

Not only did most participants view troubleshooting as a major learning goal for electronics courses, but almost all of them identified troubleshooting as a necessary skill for physics coursework, physics research, and engineering research and development. For example:
\QUOTE{[The ability to troubleshoot is] a core skill as a scientist or a physicist.}{Alder}
\QUOTE{[The ability to troubleshoot is] essential in all areas of physics. Whether you're troubleshooting electronics or your computer program. It could be your homework assignment. You could call all those things troubleshooting. It shows up everywhere in physics. Even in theoretical physics, too, it's gonna show up. I think it's an essential part.}{Yew}
Here Alder and Yew identified the ability to troubleshoot as a ``core" or ``essential" physics skill. Moreover, Yew indicated that physicists troubleshoot ``everywhere in physics," including when working on computer programs, in theoretical contexts, and on homework assignments. Troubleshooting was also identified as important for students entering professional careers other than physics research:
\QUOTE{Regardless of what kind of environment they're gonna be in---experimental laboratory or [research and development] situation---troubleshooting is a big part of the job, right? Figuring out what's wrong and how to fix it.}{Walnut}
\QUOTE{Our students who go on into industry, a lot of time it involves, `You're responsible for this system, and when this system doesn't work, what do you need to do?' In particular for students who go into small start-up companies, [the ability to troubleshoot is] crucial as well.}{Elm}
When it comes to careers in research and development, start-up companies, and industry, Walnut and Elm framed troubleshooting as a ``big" or ``crucial" aspect of work. Similarly, one participant who taught at a two-year college said that her course focuses on troubleshooting because ``most of [her students] want to become electronic technicians," implying that some instructors perceive troubleshooting to be an important skill for technician careers as well.

In previous analyses of these data, we have argued that electronics instructors' perception that troubleshooting is an important physics skill is connected to their belief that experimental systems (including electric circuits) rarely function as intended when first built~\cite{Dounas-Frazer2016b}. Additionally, some instructors articulated one or more of the following ideas when explaining why they thought it was valuable for students to learn how troubleshoot: troubleshooting circuits helps students understand how ``real" circuits and components function; students' troubleshooting ability is positively coupled to their independence and confidence; and, beyond the context of electronics, people need to troubleshoot in all areas of life, from debugging code to managing people. For example, one participant said that troubleshooting ``is broadly applicable to mechanical systems and life designs."

Despite some instructors' perception that troubleshooting was broadly relevant throughout and beyond physics, we focused our interview on the role of troubleshooting in the relatively narrow domain of electronics courses.


\subsection{Definition of troubleshooting}\label{sec:definition}

When asked to define troubleshooting in the context of electronics, almost all participants articulated at least one of the cognitive features typically associated with troubleshooting (Sec.~\ref{sec:troubleshooting}), and many identified multiple features. For example:
\QUOTE{I think of troubleshooting as the process of identifying when there could be an issue with the circuit, which is even proactive. Finding good places along building or designing a circuit to test what you have. Predict its response and verify that it meets that response. To recognize when a circuit is not behaving properly, what that might look like, and then techniques. General approaches to use to \ldots\ identify the problems in a logical manner when they exist, and then fix [the circuit].}{Alder}
Collectively, Alder and other instructors defined troubleshooting as a response to a discrepancy between the expected and actual performance of a circuit, with the goal being to get the circuit to perform as expected. Troubleshooting, according to instructors, involves isolating and fixing errors using a logical, iterative, and step-by-step process. Instructors articulated multiple features of this step-by-step process: making predictions or forming expectations, generating hypotheses or explanations, and breaking a system into smaller subsystems.

In their definitions of troubleshooting, some instructors---all of whom taught courses with final projects---distinguished between troubleshooting during the repair and design processes. For example:
\QUOTE{The troubleshooting that [students] do in [the electronics course] is not to, say, fix a broken piece of electronic equipment. Their troubleshooting is more of a test, analyze, and fix of their designs as they try to make the things work. \ldots A big part of [students] being successful on the project, I think, is to realize that they have a flawed design.}{Birch}
\QUOTE{A lot of times, we think of troubleshooting as being finding something wrong with something that's already built. But sometimes what [the students] built is fine \ldots\ The circuit is functioning, but how they anticipated how it was going to work, there were flaws in their reasoning. \ldots\ With students in particular, they can be troubleshooting [the physical circuit] when it's actually something in the design that they designed incorrectly. I think of both as being troubleshooting.}{Elm}
Birch and Elm articulated slightly different stances about whether students engage in repair as part of the electronics course. Whereas Birch said that students do not ``fix broken \ldots\ equipment," Elm implied that students do sometimes engage in identifying faults in circuits that have already been built. However, both Birch and Elm noted that one important aspect of troubleshooting in electronics courses involves revising the design of a circuit, not just its physical construction.

In electronics courses that engage students in designing and building circuits, discrepant circuit performance could be due to a design flaw in addition to a faulty component or errant connection. Indeed, design and construction considerations inform some instructors' conception of troubleshooting proficiency.


\subsection{Characteristics of proficient troubleshooting}\label{sec:proficiency}

In our study, a common description of a student who is good at troubleshooting involved three components: after encountering a problem, such a student (i) immediately engages in the troubleshooting process, (ii) only asks for help on problems that are new to the student or difficult to diagnose for both the student and the instructor, and (iii) eventually fixes the problem and gets the circuit to work. Students who are ``bad" at troubleshooting, on the other hand, ``sit on their hands and wait" for help instead of trying to diagnose or fix the problem themselves, ``immediately ask for help" even for problems they've seen before, or ``never get their work done because they don't know how" to troubleshoot. Every instructor articulated at least one of these three components, and many articulated two or three.

In addition, instructors described both cognitive and non-cognitive characteristics of troubleshooting proficiency, as well as circuit construction practices that are characteristic of competent troubleshooters. While all instructors said that they believe students can learn to troubleshoot, about half believed that some aspects of troubleshooting proficiency are innate. Below, we elaborate on instructors' conception of proficiency.


\subsubsection{Cognitive characteristics of proficiency}

{When discussing cognitive characteristics of proficiency, we focus first on types of knowledge and then on cognitive subtasks.}

Almost all of the instructors identified at least one example of strategic, domain, and/or procedural knowledge when describing troubleshooting proficiency, and many described examples of at least two of these types of knowledge.  When characterizing proficiency, many of instructors described aspects of strategic knowledge. Many of these instructors spoke generally about ``logical," ``methodological," ``organized," ``step-by-step," or ``strategic" approaches. Some instructors were more specific:
\QUOTE{The skill I'm looking for in particular is starting from where you know the signal, and moving systematically through the circuit. That's how you troubleshoot. You keep going up until the point where you lose track of what's going on.}{Juniper}
\QUOTE{Oftentimes what they don't do is, if there's two parts to a circuit that are connected together, \ldots\ they don't try to disconnect to make sure they understand on an individual basis. They try to understand the entire circuit as a whole rather than trying to break it up into small pieces.}{Yew}
Here, Juniper and Yew described the forward topographic strategy and split-half strategy, respectively. About half of instructors emphasized the importance of being able to treat complicated circuits as being comprised of multiple subsystems that can be individually tested.

Many interviewees identified aspects of domain knowledge as integral to troubleshooting proficiency. For example:
\QUOTE{If someone has a conceptual gap, they're limited as to how much they can do with the troubleshooting}{Elm}
\QUOTE{The poor troubleshooters are treating things like black boxes. Something goes in, something goes out. But the good troubleshooters have some expectation, a model. \ldots\ `According to my model, this should happen. But on my board, something else is happening.' That's pretty essential, to have an expectation.}{Tanoak}
Elm and Tanoak described students with ``limited" or ``poor" troubleshooting ability as those who have a ``conceptual gap" or who treat circuits like ``black boxes." Tanoak described ``good troubleshooters" as students who have a model that allows them to form expectations about the circuit behavior. 

About half of instructors emphasized that {procedural knowledge plays a role in overall troubleshooting proficiency}:
\QUOTE{Most of the problems [students] have early on in the term are just unfamiliarity with the test equipment.}{Birch}
Birch and other instructors identified students' lack of familiarity with oscilloscopes, multimeters, function generators, and other equipment as obstacles to successful troubleshooting.

{Many instructors included one or two cognitive subtasks} as part of their conception of troubleshooting proficiency: generating causal hypotheses and/or performing diagnostic tests. For example, when asked what they would want students to try before asking for help, Cypress and Tanoak said, 
\QUOTE{I'd love for them to think about why things caught fire before they just replace the resistor or whatever.}{Cypress}
\QUOTE{One thing I really try to get them to do is to come up with a hypothesis or a guess about what might be wrong. And then devise little tests to see if that hypothesis is correct or not.}{Tanoak}
Tanoak's desire for students to perform tests was shared by other instructors:
\QUOTE{The first thing of course is to check that they've actually assembled the circuit correctly, that the values of the components that they're using are actually the values of the components, how to read resistor color codes and all that kind of stuff to get the circuit built correctly on the breadboards that they're using.}{Birch}
\QUOTE{I really would like them to trace the path of current flow and voltage drops to make sure that every individual voltage drop or current path is basically what you would expect from a circuit analysis you did on paper.}{Dogwood}
Birch and Dogwood said, respectively, that they wanted students to perform diagnostic inspections of the circuit construction and diagnostic tests of current and voltage throughout the circuit.

{In addition to highlighting cognitive aspects of troubleshooting competence, instructors also described aspects of proficiency that we classified as non-cognitive, which we discuss in the next subsection.}


\subsubsection{Non-cognitive characteristics of proficiency}

{Many participants described non-cognitive aspects of troubleshooting proficiency, including facets of students' demeanor as well as students' ability to regulate their own emotions}. For example:
\QUOTE{The students \ldots\ who I think are weak at troubleshooting are the ones who still don't feel much confidence in trying to find their own error.}{Filbert}
\QUOTE{Troubleshooting requires attention to details. And patience. \ldots  [Students who are bad at troubleshooting] are impatient.}{Oak}
\QUOTE{They definitely work more independently, the ones who are good at troubleshooting.}{Tanoak}
According to Filbert, Oak, and Tanoak, students who are ``weak at troubleshooting" lack confidence and patience while students who are ``good at troubleshooting" work independently. Other instructors also identified confidence, patience, and/or independence as features of troubleshooting proficiency.

In addition, instructors also identified students' ability to cope with frustration or other emotional responses as necessary for competent troubleshooting:
\QUOTE{There is some level of frustration involved when something doesn't work. You've got to overcome that and do the troubleshooting.}{Maple}
\QUOTE{I would think that the [students] who are able to have a more logical response would do better at troubleshooting than the one with a more emotional response.}{Willow}
Here, Maple acknowledged that frustration is a natural reaction when a circuit doesn't work and emphasized that students nevertheless need to ``overcome" their frustration when troubleshooting. Willow, on the other hand, drew a distinction between students who have predominantly logical versus emotional responses, suggesting that the former are ``better at troubleshooting" than the latter.

Some instructors described an attitudinal aspect of troubleshooting proficiency. According to these instructors, students who are good at troubleshooting understand that circuits may not work as intended regardless of how well they were built; such students view troubleshooting is a necessary part of building circuits. For example:
\QUOTE{It's getting around this whole thing, `If I built this circuit the way the diagram says, it should work.' It's a binary thing. `I built it, it works.' You have to get over that. That's not the case. There are things that can go wrong. It's important to get over that hurdle.}{Yew}
\QUOTE{Some people think that \ldots\ if they build it and it doesn't work, obviously [they] did something wrong. But troubleshooting doesn't come to mind as part of the experimental procedure. And it should be. It should be part of the experimental procedure.}{Maple}
\QUOTE{The main point of [the final project] is for students to see how \ldots\ [troubleshooting is] gonna be very necessary because, as soon as they start making circuits that involve more than one subsystem, [they] connect them together and find they don't work. They should expect that.}{Birch}
Yew described an unproductive attitude, namely, the expectation that a circuit should work properly if it was built correctly. Yew framed this attitude as a ``hurdle" that students need to ``get over." While Yew implied that students should accept that ``there are things than can go wrong," Maple implied that students should view troubleshooting ``as part of the experimental procedure." Similarly, Birch said that students should view troubleshooting as ``necessary" and expected, especially when building complicated circuits.

In the next subsection, we describe a related idea: students who are good at troubleshooting not only expect to troubleshoot, but their anticipation of problems informs their circuit construction practices.


\subsubsection{Construction practices of proficient troubleshooters}

Students' circuits were described as looking like a ``rat's nest," like a ``spaghetti of wires," or like ``they were put together by drunk spiders." Beyond eliciting humorous analogies, students' construction practices were viewed as coupled to their troubleshooting practices. For example, some instructors noted that building neat circuits can ease the troubleshooting process:
\QUOTE{As far as troubleshooting goes \ldots\ [students] quickly learn that the style in which they do things, the way they approach the problem, the way they lay things out on the breadboard has a huge impact on their success overall. \ldots\ Slowly over time they appreciate the care they put up front plays a big role in how hard it is to troubleshoot on the back end.}{Walnut}
\QUOTE{You can probably reduce your frustration levels if \ldots\ your circuit is clean and color-coded. Your troubleshooting may become easier and so your frustration level may diminish because those techniques have been incorporated in your design.}{Maple}
Walnut said that students' construction practices not only impact their ability to build a functional circuit, but can also facilitate---or hinder---the troubleshooting process once the circuit is built. Walnut and Maple both said that careful circuit construction makes it easier to troubleshoot the circuit later. Maple further highlighted that good construction practices may mitigate frustration.

Other instructors emphasized that students should test circuits during the construction process:
\QUOTE{The mistake is building it all at once and hoping it works. More often than not it doesn't work. \ldots My impression, my gut response, is that a lot of [students] will probably try---they'll start from scratch a lot prior to receiving instruction. This breaking [the troubleshooting process] into simple steps is not something they know coming in. They'll usually scrap the whole thing and start from scratch.}{Tanoak}
\QUOTE{One thing that I hope is that [students] would be testing as they build \ldots\ If it's a multi-stage amplifier, building each stage separately and testing it before they connecting everything together.}{Alder}
According to Tanoak, some students build their circuits ``all at once" rather than troubleshooting throughout the construction process. Tanoak called this a ``mistake" because students' circuits don't typically work after the first construction attempt. When such students build a circuit that doesn't work, rather than engaging in troubleshooting, they deconstruct the circuit in its entirety and build a new one ``from scratch." In a sense, these students are resorting to reconstruction as a strategy to avoid troubleshooting altogether instead of breaking the task ``into simple steps" and engaging in the process of troubleshooting. Along these lines, Alder expressed a desire for students to build complicated circuits one subsystem at a time, and for students to test each subsystem before connecting it to other subsystems.

Thus, according to some of the instructors in our study, students who are good at troubleshooting build their circuits in ways that mitigate the difficulty of troubleshooting.


\subsubsection{Beliefs about learning how to troubleshoot}

When asked whether troubleshooting was a learnable or innate skill, all instructors said that they believe students can learn to troubleshoot. For example:
\QUOTE{It's learned, it's learned. [People] need to be taught the right way of doing it.}{Redwood}
\QUOTE{It's something people can learn. I've seen students get better at it as the semester goes on.}{Dogwood}
\QUOTE{[People] absolutely have to learn it. I don't think---there's nothing innate about that.}{Yew}
About half of instructors said that, while troubleshooting can be learned, there are nevertheless innate components to troubleshooting. In particular, some instructors expressed a belief that ``logical," ``algorithmic," or ``operational" thinking is innate. Others said that students with certain ``temperaments," ``personalities," or ``character traits" are ``intuitively predisposed" to troubleshoot well. For example, Walnut highlighted the innate nature of aggression, curiosity, and a willingness to experiment:
\QUOTE{I think some people are just naturally more willing to experiment. \ldots\ A lot of times [students are] afraid of making a mistake. A lot of times it's an innate lack of aggression, a lack of curiosity almost. And that's a worrying sign. When you see students like that, you're worried that they maybe picked the wrong career path. I don't know that you can necessarily flip that switch.}{Walnut}
When discussing whether troubleshooting can be learned, one instructor referenced a stereotype about the differences between theorists and experimentalists:
\QUOTE{I hope [troubleshooting] can be learned. I think it can be learned. I hesitate to be absolutely certain about that.  \ldots\ You know the stereotype about theoreticians who don't know what end of the screwdriver to hold on to. You see that in the lab. There are geniuses at the chalkboard solving quantum problems, but you shudder when they come into the lab.}{Cypress}
Cypress simultaneously expressed hope, belief, and uncertainty that troubleshooting can be learned. To explicate his uncertainty, Cypress called upon a stereotypical dichotomy between theoretical and experimental competence: the ``geniuses at the chalkboard" who ``don't know what end of the screwdriver to hold on to." This stereotype is consistent with the idea that some students more than other are naturally inclined to be good at troubleshooting electric circuits.


\subsection{Teaching practices}\label{sec:teaching}

In our study, the most prevalent form of direct instruction was through interactions between the instructor and the lab group, which consisted of one, two, or three students:
\QUOTE{{One of the things about teaching troubleshooting that I've found hard is that I have not found a good way to just do it. It's a lot of having conversations with students as they're working.}}{Elm}
Instructors also said that they addressed troubleshooting in lectures and lab manuals, often in the form of highlighting common pitfalls for a particular lab activity. Few instructors described having labs specifically dedicated to improving students' ability to troubleshoot; instead, instructors typically viewed all lab activities as opportunities for students to practice that skill. In this subsection, we elaborate on these teaching practices.


\subsubsection{Practices consistent with cognitive apprenticeship}

Instructors in our study described teaching about troubleshooting via interactions with small groups of students. These interactions were often in response to students encountering a problem with their lab activity. For example:
\QUOTE{I'm actually wandering around looking over [students'] shoulders. I spend 5 to 10 minutes just standing behind somebody watching. If they're doing fine, I'll go away. So I'm always there. They typically don't have to call me over. I've got my [undergraduate learning assistants] trained to do that also. They're moving around watching someone do something.}{Juniper}
\QUOTE{That's one of the fun parts of electronics is that there's hands shooting up all over the place, which I think is really great especially because the class isn't that large. There's a lot of one-on-one instruction. \ldots\ Physics, as much as it can be, is apprenticeship. Working alongside people.}{Tanoak}
Juniper described a learning environment in which the instructor and teaching assistants are ``wandering around" the classroom, ``watching" students, and intervening when necessary; students ``typically don't have to call" for help because Juniper is ``always there." Tanoak described a slightly different situation: in Tanoak's case, students raise their hands to ask for help. According to Tanoak, students still receive ``one-on-one instruction," which Tanoak characterized as ``apprenticeship."

When helping a lab group troubleshoot their circuit, almost all instructors said they employed one or more of the following practices: asking questions, coaching, verbalizing reasoning, modeling how to troubleshoot, and fading support. Many participants described using two or more of these practices, and some described using three or more of these practices.

For example, Elm described asking questions of students in order to get students to articulate their reasoning:
\QUOTE{I try to probe them. `What do you expect and why do you think this isn't what you expect?' \ldots\ I am almost never going to help them right away. I will occasionally push them with questions so they can think about what they might want to think about. But even then I won't say, `Have you thought about this?' I ask, `What do you think might be wrong? Where might you be having problems?'}{Elm}
To answer Elm's questions, a student would need to have formed an expectation or hypothesis and be able to verbalize it. According to Elm, these questions are often provided in lieu of direct help. The goal of asking students such questions is explicitly \emph{not} to seed ideas about potential problems, but rather to get students to ``think about what they might want to think about." In contrast to Elm, Sycamore provided students with more direct guidance:
\QUOTE{If I don't see an obvious solution, then I would tell them what to check for. Check for this current, check for that LED, and so on.'}{Sycamore}
Here Sycamore described coaching students about ``what to check for" when they encountered difficult problems with their circuits.  Many instructors asked questions of students in a manner similar to Elm, and about half of the instructors coached their students by making suggestions similar to those outlined by Sycamore.

Verbalizing reasoning and modeling how to troubleshoot were each practiced by some instructors. For example, when students ask for help with problems, Alder and Elm said,
\QUOTE{I definitely try to talk out loud. I don't know if I'm speaking in as great of detail as I think I am. But I am trying to talk through my process with them. \ldots\ Talking through the process of how you determine that the op-amp itself seems to be not functioning.}{Alder}
\QUOTE{When I get to their board, to their station, and they're stumped, I'll sit down and go through my systematic approach to troubleshooting. I'll check power, check that the board's plugged in. All those things. Then I'll look at the wiring. Then I'll systematically check, poking around with the [oscilloscope], at various places in the circuit. They see us doing this as freshman and mimic our approach.}{Walnut}
Both instructors said that they troubleshot students' circuit themselves, but in slightly differing ways. Alder described talking aloud while troubleshooting the circuit, thus verbalizing ``the process of how you determine" the fault in a given circuit. Elm, on the other hand, described demonstrating the troubleshooting process so that students can observe---and later mimic---practices like ``check[ing] power," ``look[ing] at the wiring," and ``poking around with the [oscilloscope]."

Some instructors described fading their support over the course of the semester. For example:
\QUOTE{I try to scale the help that I give. They know now that, if they call me over, the first question I'll ask is, `Well, what did you measure? Have you tried to work your way through the circuit?' This wasn't my first question early in the semester. \ldots\  I do ask them to show me what they've done at this point. But two or three weeks ago, I would say, `This is the way I would do it. Get your meter, make sure you have a common ground, and check different spots.'}{Filbert}
Here, Filbert said she scales her support in the following way. Early in the semester, she coaches students about which measurements they should be making. As the semester progresses, she stops coaching students and starts asking them ``to show [her] what they've done" by asking questions like, ``What did you measure?" For Filbert, fading (or scaling) support involves transitioning away from coaching students and towards asking them to verbalize their process.

In addition to engaging in apprenticeship-style interactions with students during lab, instructors in our study also described teaching students how to troubleshoot by specifically addressing troubleshooting in their lectures, activities, and/or course materials.


\subsubsection{Lectures, lab guides, and activities}

Although some instructors said they do not specifically address troubleshooting in lectures, many instructors described lectures or whole-class discussions focused on troubleshooting. Lectures and discussions about troubleshooting were typically described as informal or impromptu, with the goal of highlighting troubleshooting tips or common pitfalls relevant to a particular lab activity. For example:
\QUOTE{{The beginning of each lab starts with a 40-minute pre-lab lecture/question section where you talk about the specifics of that particular lab, things to look for, problems to look out for, and those sorts of things. I think that's probably where, since each lab has it's own content, where we talk about the troubleshooting for that particular lab.}}{Pine}
\QUOTE{I have traditionally done a very short lecture at the beginning of each class to get the [students] together and get focused on what was being done. And oftentimes I'll talk about, we'll have a group discussion about, a common problem. I haven't done it very formally. I haven't really talked through the process formally. We'll have these informal discussions and I'll give pieces of advice. A lot of it is as things come up.}{Tanoak}
\QUOTE{{It's not formal teaching.  We don't have that formally integrated into the curriculum. It's more informal in the lab environment. \ldots\ For example, we are building a circuit and we are working with that and students ask some questions. \ldots\ At that time you can, for example, talk to them about other possibilities or issues we have seen in the past.}}{Holly}
Here, Pine, Tanoak, and Holly said that their lectures and discussions focused on ``the specifics of a particular lab," ``a common problem," and ``issues [instructors] have seen in the past." Whereas Pine said that such lectures happen at the beginning of lab, Tanoak and Holly described their discussions as ``informal" and said that discussions were responses to problems that ``come up" or to ``students ask[ing] some questions" about the circuit they are working on. Despite being informal, impromptu, {and/or tailored to the details of a particular lab activity}, instructors typically described such lectures and discussions as happening regularly throughout the semester.

About half of instructors said they do not specifically address troubleshooting in syllabi, lab manuals, handouts, or other curricular materials. The other half of instructors said they did address troubleshooting in their curricular materials. Like the troubleshooting-oriented lectures and discussions, these materials were typically described as focusing on general tips and common pitfalls relevant to a particular lab activity. For example:
\QUOTE{{The introductory lab manual deals with [troubleshooting] a fair amount. So, mostly in the first few labs and in the introductory material up front. The introduction to the whole lab manual we discuss [troubleshooting]. And the first few labs deal with [troubleshooting] and then it pops up as these pitfall things.}}{Walnut}
\QUOTE{Some of the lab handouts have tips. For example, don't measure resistance while the resistor is in the circuit. But probably not in more detail than that. Don't use an ammeter like a voltmeter. Minor things.}{Dogwood}
{Walnut and Dogwood said that their lab manuals addressed troubleshooting by highlighting ``pitfalls" and ``tips." Each instructor described using lab manuals that address troubleshooting in different levels of detail. Whereas the manual in use by Walnut addresses troubleshooting in the introduction, the manuals in use by Dogwood focus on ``minor things" like proper use of measurement devices. Redwood and Maple also said they addressed troubleshooting in their lab manuals:}
\QUOTE{{You mean the lab guides? Oh, yes. General rules about grounding, electric shocks, resistors, voltages. General, but not specific.}}{Redwood}
\QUOTE{{Some labs indicate that you may want to keep your wiring neat. Color coding is important. Write your pinouts and chip numbers and things like that.}}{Maple}
{Redwood and Maple described manuals that focus on ``general rules" and advice for constructing circuits that are properly grounded and color-coded. Beyond giving tips, rules, and advice about measuring and constructing circuits, instructors in our study did not describe lab manuals that provided students with strategic information about how to troubleshoot a circuit.}

While some instructors said they have implemented lab activities that were specifically designed to improve students' troubleshooting ability, many said they had not. About half of instructors said there is no need for such activities since the need to troubleshoot arises naturally in every lab activity, as we have discussed in more detail elsewhere~\cite{Dounas-Frazer2016b}. Additionally, some instructors said that they don't have time to dedicate a lab solely to troubleshooting.

Among those instructors who had designed troubleshooting-oriented activities, two types of activity were described: those that engage students in the repair of a malfunctioning circuit and those that ask students to analyze the behavior of circuit schematics that were deliberately drawn with faults in them. A few instructors said they included troubleshooting-oriented questions of the latter type in pre-lab homework assignments. 

Many instructors expressed a desire for research-based lab activities, tutorials, and/or worksheets specifically focused on troubleshooting. Some instructors said they also wanted a handout summarizing common troubleshooting strategies for electric circuits.


\subsection{Assessment practices}

When asked whether they had designed an activity to test students' troubleshooting skills, many participants said they had not done so. Of those who said they did not assess troubleshooting ability, some implied that the skill was assessed indirectly:
\QUOTE{Most of the labs that they do are gonna test [troubleshooting ability] to some extent, but I wouldn't say there's something specifically designed to test that.}{Pine}
\QUOTE{Not explicitly, other than what are in the labs already. \ldots\ We always have [troubleshooting] in the back of our minds, that it's one of the key things [students are] getting out of the course. Implicitly, but not explicitly.}{Evergreen}
According to Pine and Evergreen, most lab activities facilitate assessment of students' troubleshooting ability ``to some extent" or ``implicitly" rather than ``specifically" or ``explicitly." As we have argued elsewhere~\cite{Dounas-Frazer2016b}, this perception is coupled to the belief that the need to troubleshoot arises naturally on most activities, and hence successful circuit construction can be used as a proxy for troubleshooting competence.

About half of participants provided an example of a troubleshooting assessment. Four types of assessment were described: those that involved talking with students, repairing a malfunctioning circuit, testing students' familiarity with test and measurement equipment, {or asking students to document problems with their final projects}. For example:
\QUOTE{You can usually tell [if a student is good at troubleshooting] in the conversation. \ldots Another thing I love about electronics is that there are like all of these oral exams all the time. I think oral exams are the best thing ever. You have this conversation with the student, and you can very quickly assess that they don't understand. You get right to the part where they have no clue what's going on. It's in these conversations that you quickly get a picture of their troubleshooting.}{Tanoak}
\QUOTE{We put the faulty element in the circuit and then we ask [the students], `What is the problem?' We put the wrong component with the wrong resistivity or capacitance. \ldots This is part of the test at the end of the semester. We intentionally ask them to troubleshoot.}{Redwood}
\QUOTE{One [test] was trying to get an oscilloscope to work. I would hook it up wrong to start with. Most common way is to not have something grounded. Hook a single probe up to it and say [to the student], `Here, make this work.'}{Cypress}
\QUOTE{{As part of their final [microcontroller] project, one of the written components they have to turn in I refer to as `troubleshooting notes.' \ldots\ I'm looking for, at that point, this intentional approach to identifying challenges or looking for places to stop and look for bad behavior in their circuit. But that's the only way I've figure out so far, is to basically ask them to turn in a written document that actually goes through their troubleshooting approach.}}{Alder}
{As opposed to indirect assessments of troubleshooting that use successful circuit construction as a proxy, Tanoak, Redwood, Cypress, and Alder described more direct assessments of students' troubleshooting ability. However, in our dataset, such assessments of troubleshooting competence were highly idiosyncratic and no obvious patterns emerged with respect to direct assessment of students' troubleshooting ability.} Some instructors expressed an interest in using research-based troubleshooting assessments in their classrooms.


\section{Summary and Discussion}\label{sec:discussion}

{We designed an exploratory qualitative study in which instructors were interviewed about their practices  related to troubleshooting instruction in electronics lab courses. Audio data were collect for 20 instructors from 18 institutions that varied in terms of size, selectivity, and student populations. We characterized instructors' perceptions about the role of troubleshooting in electronics courses and physics research, their conceptions of troubleshooting and troubleshooting proficiency, and their practices for teaching and assessing troubleshooting.}

{Our results provide insight into instructional practices related to troubleshooting electric circuits. We focus on {four} major findings of this work, and we couple those findings to our research questions RQ1--RQ6. In some cases, our findings give rise to additional research questions, which we briefly describe where appropriate.}

First, we found that almost all instructors in our study said that developing students' ability to troubleshoot was a central learning goal for electronics courses (RQ1) and that troubleshooting is crucial skill for physicists, engineers, and people who want to pursue careers in research and development (RQ2). This result complicates the findings of Coppens et al.~\cite{Coppens2016a}, who asked instructors  to rank the importance of potential learning goals for electronics lab courses from a list of goals that did not include an option related to troubleshooting. Although instructors in our study identified other learning goals in addition to troubleshooting---including some that align with the findings of Coppens et al.~\cite{Coppens2016a}, such as learning how to use test and measurement equipment---our work does not provide insight into the relative importance of troubleshooting compared to other potential learning goals. Further investigations would be needed to comprehensively characterize instructors' learning goals for electronics lab courses.

{Second, we found that almost all instructors defined troubleshooting according to the cognitive subtasks typically associated with troubleshooting (RQ3) and described a multifaceted conception of troubleshooting proficiency (RQ4). Some instructors noted that, in an electronics course, troubleshooting goes beyond repairing a formerly functional or poorly built circuit. According to these instructors, students often make mistakes with their circuit designs, and hence students may also need to revise their designs during the troubleshooting process. In addition to identifying cognitive aspects of troubleshooting competence---including mastery of multiple types of knowledge, cognitive subtasks, and strategies---instructors identified confidence, patience, independence, emotional regulation, and attitude as hallmarks of proficiency. This finding builds on other work that highlights connections between troubleshooting and students' levels of confidence, patience, and frustration~\cite{MacPherson1998,Estrada2012}.}

{Third, among instructors in our study, the predominant form of explicit instruction about troubleshooting aligned with the cognitive apprenticeship paradigm of instruction (RQ5). Instructors said they interact with one, two, or three students at a time as problems arise during lab activities. In response to such problems, almost all instructors described engaging in articulation, coaching, modeling, and/or fading support. Lectures on troubleshooting were less commonly used by instructors in our study, and were typically characterized as impromptu or informal and focusing on specific tips and hints for a particular activity, not on troubleshooting strategies or subtasks more generally.} However, this study does not provide insight into instructors' underlying beliefs about how students learn. Farnham-Diggory~\cite{Farnham-Diggory1994} argued that apprenticeship is distinguished from other models of teaching not by practices like coaching or modeling, but by conceptions about how novices develop expertise: in the apprenticeship model, novices are thought of as sociologically different from experts and expertise is achieved through acculturation. Further research is needed in order to probe whether and how electronics lab instructors' beliefs about learning align with their apprenticeship-style teaching practices.

{Fourth} and lastly, while the instructors in our study identified troubleshooting as an important learning goal for their electronics course, about half of them said they did not assess students' troubleshooting ability and, among those instructors who did describe a troubleshooting assessment, no instructors described comprehensive, scalable assessments (RQ6). In addition, instructors noted that proficient troubleshooting includes anticipating the need to troubleshoot and hence building neat, color-coded circuits; meanwhile, instructors identified meticulous construction practices as a barrier to using students' ability to build a working circuit as a proxy for troubleshooting ability. Therefore, there is a need for development of research-based, process-oriented assessments that focus on students' ability to troubleshoot electric circuits.

However, improving students' troubleshooting ability is not the only learning goal for electronics courses; indeed, instructors in our study articulated multiple goals---including the ability to model circuits. In previous work~\cite{Dounas-Frazer2016a}, we demonstrated that troubleshooting and model-based reasoning were complementary and mutually reinforcing practices. Additional research on electronics courses may help clarify the appropriate scope of research-based assessments for electronics courses. {In the future, we aim to develop research-based assessments for use in upper-division electronics lab courses. This work highlights the need for such assessments to focus, at least in part, on students' ability to troubleshoot. Ultimately, assessments of student learning in instructional lab environments will pave the way for the development and implementation of evidence-based curricular transformations in upper-division lab courses.}


\begin{acknowledgements}
{The authors acknowledge the PER group at CU Boulder generally, and Bethany Wilcox in particular, for useful input on study design and interpretation of results.} This material is based upon work supported by the NSF under Grant Nos. DUE-1323101 and PHY-1125844.
\end{acknowledgements}


\bibliography{ts_database_160701}

\end{document}